%Paper: hep-ph/9311270
%From: GENSINI@vaxpg.pg.infn.it
%Date: Thu, 11 Nov 1993 11:10:12 +0100 (WET)

\magnification=1200
\parindent 1.0truecm
\rm
\null

% frontespizio

\footline={\hfil}
\vglue 0.8truecm
\rightline{\bf DFUPG-66-93}
\rightline{\sl September 1993}
\vglue 2.0truecm
\centerline{\bf SU(3)--Breaking Effects in Axial--Vector Couplings of Octet
Baryons }
\vglue 1.0truecm
\centerline{ Paolo M. Gensini}
\centerline{\sl Dip. di Fisica dell'Universit\`a di Perugia, Perugia,
Italy, and }
\centerline{\sl Sezione di Perugia dell'I.N.F.N., Perugia, Italy}
\centerline{ and }
\centerline{ Galileo Violini }
\centerline{\sl Dip. di Fisica dell\'\ Universit\`a della Calabria, Rende
(Cosenza), Italy, }
\centerline{\sl I.N.F.N., Gruppo Collegato di Cosenza, Italy, and }
\centerline{\sl Centro Internacional de F\'\i sica, Santa Fe de Bogot\'a,
Colombia. }
\vglue 3.0truecm
\centerline{Talk presented at}
\centerline {{\sl 5th Int. Sympos. on Meson--Nucleon Physics and the Structure
of the Nucleon, }}
\centerline{{\sl Boulder, CO, sept. 1993}}
\vglue 2.0truecm
\centerline{ To be published in }
\centerline{\sl $\pi N$ Newsletter }
\pageno=0
\vfill
\eject
% fine del frontespizio

\footline={\hss\tenrm\folio\hss}
\centerline{\bf SU(3)--Breaking Effects in Axial--Vector Couplings of Octet
Baryons }
\vglue 0.5truecm
\centerline{ Paolo M. Gensini}
\centerline{\sl Dip. di Fisica dell'Universit\`a di Perugia, Perugia,
Italy, and }
\centerline{\sl Sezione di Perugia dell'I.N.F.N., Perugia, Italy}
\centerline{ and }
\centerline{ Galileo Violini }
\centerline{\sl Dip. di Fisica dell\'\ Universit\`a della Calabria, Rende
(Cosenza), Italy, }
\centerline{\sl I.N.F.N., Gruppo Collegato di Cosenza, Italy, and }
\centerline{\sl Centro Internacional de F\'\i sica, Santa Fe de Bogot\'a,
Colombia. }
\vglue 0.5truecm

\centerline{\bf ABSTRACT }

\par Present evidence on baryon axial--vector couplings is reviewed, the main
emphasis being on internal consistency between asymmetry and rate data. A
complete account of all {\sl small} terms in the Standard Model description of
these latter leads to {\sl both} consistency {\sl and} evidence for breaking
of flavour SU(3) in the axial couplings of octet baryons.
\vglue 0.5truecm

\leftline{\bf 1. Introduction. }

\par The results we are going to present here constitute a preliminary to a
complete study of $S = - 1$ systems, and in particular of the low--energy,
coupled $\bar KN$, $\pi\Sigma$, $\pi\Lambda$ channels. We shall concentrate
on the description of unpolarized--baryon semi--leptonic decay rates, and
in particular on the internal consistency (recently questioned by Jaffe and
Manohar$^1$) of the present data sets (PDG averages) for both asymmetries and
rates.

\par As a conclusion, we shall present
\item{a)} {A new determination of $\vert V_{ud}\vert$ and $\vert V_{us}\vert$;}
\item{b)} {A set of values for the {\sl measured} axial couplings;}
\item{c)} {Evidence for breaking of flavour SU(3) symmetry in the latter.}

\par Despite recent experimental progress$^2$, and the fact that a theoretical
analysis of these data is standard business$^3$, one still finds in the
literature statements which are at best confusing, and sometimes blatantly
wrong. For instance, as late as 1992 a review paper$^4$ is still quoting the
analysis by Bourquin {\sl et al.}$^5$ as the {\sl state--of--the--art} for
octet--baryon $\beta$ decays.

\par Use of flavour SU(3) $F$, $D$ constants for the axial charges is still
common practice, without or with only handwaving estimates of systematics,
often quoting again Bourquin {\sl et al.} as supporting evidence! It is high
time to warn that their work$^5$, though oustanding for its days, has been
(almost) completely superseded by the new evidence$^2$, and that use of their
$F$, $D$ values is no longer advisable, since they chose the {\sl wrong} value
for $g_A$, the nucleon axial coupling.
\vglue 0.5truecm

\leftline{\bf 2. The formalism.}

\par In describing the decays $A\to B\ell\nu$ for unpolarized initial and
final baryons, we use the notations $\Sigma = M_A + M_B$ and $\Delta = M_A -
M_B$, express the differential rates in terms of the variable $q^2$ (where $q$
is the lepton--pair four--momentum), and the total rates in terms of the
adimensional ratios $\delta = \Delta/\Sigma$ and $x = m_\ell/\Delta$.

\par The vector and axial--vector weak transition currents are fully
decomposed$^3$ via the real--analytic form factors $F_i(s)$ and $G_i(s)$ ($i =
1, 2, 3$) as
$$
V_\mu(q) = \gamma_\mu \cdot F_1(q^2) + i \sigma_{\mu\nu} q^\nu \cdot F_2(q^2)
+ q_\mu \cdot F_3(q^2)
\eqno (1)
$$
and
$$
A_\mu(q) = \gamma_\mu \gamma_5 \cdot G_1(q^2) + q_\mu \gamma_5 \cdot G_2(q^2)
+ i \sigma_{\mu\nu} \gamma_5 q^\nu \cdot G_3(q^2) \ ,
\eqno (1')
$$
where $F_3$ and $G_3$ represent second--class terms and vanish in the
flavour--symmetry limit. Since we choose {\sl ab initio} to work in the real
world of broken flavour symmetry, their contributions have to be numerically
checked (case by case), and we keep them throughout, differing in this from
recent, similar works$^6$.

\par Integrating over the leptons' kinematical variables we obtain the
differential rate (to be found, plus a few annoying misprints, at pages 44 and
156 of Pietschmann's handbook$^3$), to which one must add weak and e.m.
radiative (and Coulomb) corrections before comparing with the data. The final
answer can be cast into the simple expression, after inclusion of all
electroweak corrections and integration over the $q^2$--variable,
$$
\Gamma(A\to B\ell\nu) = {{G_F^2 \vert \tilde V_{KM} \vert^2}\over{60\pi^3}}
\cdot ({{\Sigma}\over{2 M_A}})^3 \cdot \Delta^5 \cdot [ F_V^C(\delta,x) +
\gamma^2 \cdot F_A^C(\delta,x) ] =
$$
$$
= \Gamma_0 \cdot [ F_V^C(\delta,x) + \gamma^2 \cdot F_A^C(\delta,x) ] \ ,
\eqno (2)
$$
where $\gamma = G_1(0)/F_1(0)$ for all cases but $\Sigma\to\Lambda e\nu$
transitions, where we define $\gamma = \sqrt{3/2} \cdot G_1(0)$, to reduce its
SU(3)--limit value to the constant $D$. The short--range, electroweak
radiative corrections are included defining$^7$
$$
\tilde V_{KM} = V_{KM} \cdot (1 + \delta_W)^{1/2} \ ,
\eqno (3)
$$
with $\delta_W = 0.0122(4)$ as given by Woolcock$^8$; besides, we write$^9$
$$
F_{V,A}^C(\delta,x) = F_{V,A}(\delta,x) \cdot (1 + \delta_\alpha) + \delta
F_{V,A}^C(\delta,x) \ ,
\eqno (4)
$$
with $\delta_\alpha$ the radiative correction, and $\delta F_{V,A}^C$ the
Coulomb correction to be included for a charged $B$, and the uncorrected
$F_{V,A}$ come from integrating the differential rates over $q^2$.

\par In the limit $\delta\to 0$, eq. (3) reduces to $\Gamma = \Gamma_0 \cdot
r_V^C(x) \cdot [ 1 + 3 \gamma^2 ]$, good to describe the neutron $\beta$
decay, but leading to inaccurate results if used for the hyperon $\beta$ (and
muonic) decays. A description of these decays, which aims at both reproducing
accurately the data and investigating size and structure of possible SU(3)
breaking in their axial couplings, must necessarily account for {\sl all}
small terms depending on $\delta$ in eq. (2), besides the obvious kinematics
coming from traces over $\gamma$--matrices. Indeed, the latter turn out {\sl
not} to dominate these {\sl kinematical} symmetry--breaking effects.

\par The first of these effects (not always treated consitently in the
literature) is the momentum dependence of the form factors. For electric-- and
magnetic--type, vector and axial--vector form factors $F_{1,2}(s)$ and
$G_{1,3}(s)$ (though the last is second--class, the approximation is supported
by $m(a_1)\simeq m(b_1)$ and the strong mixing of the $Q$--states in the
lowest, $J^{PC}=1^{+\pm}$ SU(3) meson multiplets), we assume the dipole forms
$$
F_1(s)/F_1(0) = F_2(s)/F_2(0) = ( 1 - {s\over m_V^2})^{-2}
\eqno (5)
$$
and
$$
G_1(s)/G_1(0) = G_3(s)/G_3(0) = ( 1 - {s\over m_A^2})^{-2} \ ,
\eqno (5')
$$
with $m_V(\Delta S=1)^2-m_V(\Delta S=0)^2=m_A(\Delta S=1)^2-m_A(\Delta S=0)^2=
m(K^*)^2 - m(\rho)^2$ to ensure the correct variation in the mass scales with
the $\Delta S$ of the weak currents, and $m_{V,A}(\Delta S=0)$ are taken from
the most recent fits to $ep\to ep$, $\bar\nu p\to\ell^+n$ and $\nu n\to\ell^-
p$ processes.

\par With these dipole form for $G_1$ ($F_1$), we describe the pseudoscalar
(scalar) form factors $G_2$ ($F_3$) with the only inputs of the masses of the
pseudoscalar (scalar) states, which we assume to dominate the divergences of
the axial--vector (vector) currents. Writing an ``extended'' PCAC identity
like
$$
\Sigma \cdot G_1(s) + s G_2(s) = {{\sqrt{2} f_P g_{PAB}}\over{1 - s/m_P^2}} +
{{\Delta_{GT}}\over{1 - s/m_{P'}^2}}
\eqno (6)
$$
(which continues to $s \neq 0$ the Goldberger--Treiman relation [GTR]), and
taking advantage of the {\sl numerical} equality (within errors, quite large
in the $\Delta S = 1$ case) for the GTR discrepancies $\Delta_{GT}$, $(1 - 2
m_P^2/m_A^2) \cdot [1 - \Delta_{GT}/(\Sigma G_1(0))]\simeq 1$, we rewrite the
pseudoscalar form factors $G_2(s)$ as
$$
G_2(s) \simeq {{\Sigma G_1(0)}\over{m_P^2}} \cdot [ (1 -
{{2m_P^2}\over{m_A^2}} + {{m_P^2 s}\over{m_A^4}}) (1 - {s\over m_A^2})^{-2} +
{{2 m_P^2}\over{m_A^2}} (1 - {s\over m_{P'}^2})^{-1} ] \cdot (1 - {s\over
m_P^2})^{-1} \ ,
\eqno (7)
$$
with the masses $m_P$, $m_{P'}$ taken from the PDG tables$^2$ for both
$\Delta S=0$ ($\pi$, $\pi'$) and $1$ ($K$, $K'$) transitions.

\par In exactly the same way we use the scalar analogue to the GTR
$$
\Delta \cdot F_1(s) + s F_3(s) \simeq {{\Delta F_1(0)}\over{1 - s/m_S^2}}
\eqno (8)
$$
(but for the $\Sigma\to\Lambda e\nu$ case, when $F_1$ vanishes identically),
and we obtain
$$
F_3(s) \simeq - {{\Delta F_1(0)}\over m_V^2} \cdot (2 - {m_V^2\over m_S^2} -
{s\over m_V^2}) \cdot (1 - {s\over m_V^2})^{-2} (1 - {s\over m_S^2})^{-1} \ ,
\eqno (9)
$$
still taking the masses $m_S$ (respectively of the $a_0$ and $K_0$ mesons)
from the PDG tables$^2$. To preserve the scaling with $\Delta/m_V$ of the
(second--class) scalar form factor, we use for the $\Sigma\to\Lambda e\nu$
transitions $F_3(0) \simeq - (\Delta/m_V^2) (2 - m_V^2/m_S^2) \xi$, with the
above $s$--dependence, and take as a reasonable {\sl ansatz} $\vert\xi\vert
\simeq1$.

\par One can not fix in this way the size of the (second--class) pseudotensor
form factor $G_3(0)$, for which we assume a dipole behaviour away from $s =
0$; detailed models of the baryons' wavefunctions are needed to give
predictions for such a parameter. Indeed, all models based on an SU(6) type of
quark wavefunction tend to give small values$^{10}$, and (to the best of our
knowledge) no predictions are available from other models. To ensure the
symmetry limit, we have chosen to parametrise $G_3(0)$ as $G_3(0) = - \ \Delta
\ G_1(0) \ \rho/\Sigma^2$, throwing all our ignorance in the scale parameter
$\rho$, hopefully such that $\vert\rho\vert\le O(1)$. Since in the
following we shall use $\rho = 0$, one might wonder why include this term at
all: we have two motivations, the first that one has always to gauge the {\sl
theoretical} systematics (different for $\rho$ in the rates and in the
asymmetries), and the second that consistency requires to consider {\sl
all} terms vanishing (as $\Delta^2$) in the symmetry limit.
\vglue 0.5truecm

\leftline{\bf 3. Technical remarks and analysis of data.}

\par We turn now to the steps in which we integrate the differential rates and
add the Coulomb and radiative corrections: the two steps are {\sl not}
independent, since Coulomb corrections are easier to express in tems of a
power series in the maximum recoil energy$^9$ (due to the simplifications
occurring in the static limit) $T_R = \Delta^2\cdot(1 - x^2)/(2 M_A)$, and
thus as a power series in $\delta^2\cdot(1-x^2)$. It is therefore practical to
expand the differential rates in powers of $\delta$ and then integrate term by
term, because the resulting expansion coincides with the analogous one for the
Coulomb corrections calculated developing in $T_R/M_A$ around the static limit.

\par We write the {\sl uncorrected} functions in eq. (4) as
$$
F_V(\delta,x) = F_1(0)^2 \cdot [ \phi_1(\delta,x) + 2 \kappa^2 \delta^2
\phi_2(\delta,x) + 6 \kappa \delta^2 \phi_3(\delta,x) +
$$
$$
+ {3\over2} ( 2 - {m_V^2\over m_S^2})^2 \delta_V^4 x^2 \phi_4(\delta,x) + 3\
(2 - {m_V^2\over m_S^2}) \delta_V^2 x \phi_5(\delta,x) ] \ ,
\eqno (10)
$$
which for $\Sigma\to\Lambda e\nu$ transitions reduces to
$$
F_V(\delta,x) = 2 (\Sigma F_2(0))^2 \delta^2 \phi_2(\delta,x) + {3\over2} (2 -
{m_V^2\over m_S^2})^2 \delta_V^4 x^2 \xi^2 \phi_4(\delta,x) \ ,
\eqno (10')
$$
and
$$
F_A(\delta,x) = 3\ F_1(0)^2 \cdot [ \chi_1(\delta,x) + {3\over2} \delta_P^4
x^2 \chi_2(\delta,x) + 3 \delta_P^2 x^2 \chi_3(\delta,x) +
$$
$$
+ 2 \rho^2 \delta^4
\chi_4(\delta,x) + 6 \rho \delta^2 \chi_5(\delta,x) ] \ ,
\eqno (11)
$$
where (for all cases but $\Sigma\to\Lambda e\nu$, where $\Sigma F_2(0) =
\sqrt{2} \mu_{\Sigma\Lambda}$) $\kappa = \Sigma F_2(0)/F_1(0)$ are the SU(3)
extensions of the (isovector) magnetic moment used for neutron $\beta$ decay,
and $\delta_V = \Delta/m_V$, $\delta_P = \Delta/m_P$.

\par To avoid an exponential increase of the coefficients in the series with
their indices $n$, the expansion variable for $\phi_k$ and $\chi_k$ must {\sl
not} be $\delta$, as sometimes stated to justify neglet of some or all the
previous, small corrections. One has rather to choose $\phi_k =
\sum_{n=0}^\infty f_k^{(n)}(x) \delta_V^{2n}$ (for all $k$) and $\chi_k =
\sum_{n=0}^\infty g_k^{(n)}(x) \delta_k^{2n}$, with $\delta_k = \delta_A =
\Delta/m_A$ for $k =$ 1, 4 and 5, while $\delta_k = \delta_P$ instead (and
thus much larger than the above two) for $k =$ 2 and 3.

\par To reach high accuracy without too many terms in the series, we use
``accelerated convergence'': in simpler terms, we substitute the power series
with continued fractions and truncate the latter rather than the former. A
third--order approximation turns out more than adequate for the precision
required, and we have, for instance,
$$
\phi_k(\delta,x) = {{ f_k^{(0)}(x) f_k^{(1)}(x) - \delta_V^2 [ f_k^{(0)}(x)
f_k^{(2)}(x) - f_k^{(1)}(x)^2 ]}\over{f_k^{(1)}(x) - \delta_V^2 f_k^{(2)}(x)}}
\ .
\eqno (12)
$$

\par We lack here the space to tabulate all the coefficients $f_k^{(n)}(x)$ and
$g_k^{(n)}(x)$, and we refer the readers to the full version of this
work$^{11}$. Neither shall we describe here the techniques for the electroweak
corrections terms in eqs. (3,4): again, readers are referred to the details
contained in the original papers$^{7,8,9}$.

\par Using the {\sl corrected} values for $F_{V,A}^C(\delta,x)$, and the data
listed in the PDG tables$^2$ for both the rates $\Gamma(A\to B\ell\nu)$ and
the ratios $\gamma = G_1(0)/F_1(0)$ obtained from the asymmetries, one can
follow two paths in the use of eq. (3): {\sl a)} extract the CKM matrix
elements $\vert V_{ud}\vert$, $\vert V_{us}\vert$ from the {\sl experimental}
values of $\Gamma$ {\sl and} $\gamma$, or {\sl b)} use CKM unitarity, the
bounds on $\vert V_{ub}\vert$ from charmless $B$--meson decays and the value
for $\vert V_{ud}\vert$ from superallowed nuclear $\beta$ decays (assuming
only three families of ``light'' quarks), to extract the absolute values
$\vert\gamma\vert$ from the rates $\Gamma$. In both cases we have to invoke
both the Ademollo--Gatto theorem$^{12}$ for the vector {\sl charges} $F_1(0)$
(note that arguments about its violation do not always separate the {\sl
charges} from the {\sl total vector couplings}, a dangerous attitude when
working with models for baryon states {\sl at rest}), and the approximation
$\rho\simeq0$ (to make connections with the asymmetry values for $\gamma$, all
derived under this assumption).
\vglue 0.6truecm

\centerline{\bf Table I}
\centerline{\bf Moduli of the KM--matrix elements }
\vglue 0.3truecm
\hrule
$$\vbox{\halign{#\hfil&\qquad\hfil#\hfil&\qquad\hfil#\hfil\cr
\qquad Decay & from data & recommended by PDG \cr
$\Delta S = 0$: & & \cr
$n \to p e \bar\nu_e$ & 0.97310 $\pm$ 0.00213 & 0.9747 -- 0.9759 \cr
$\Delta S = 1$: & & \cr
$\Lambda \to p e \bar\nu_e$ & 0.22478 $\pm$ 0.00348 & \cr
$\Lambda \to p \mu^- \bar\nu_\mu$ & 0.2305 $\pm$ 0.0259 & \cr
$\Sigma^- \to n e \bar\nu_e$ & 0.22308 $\pm$ 0.00480 & \cr
$\Sigma^- \to n \mu^- \bar\nu_\mu$ & 0.21009 $\pm$ 0.00972 & \cr
$\Xi^- \to \Lambda e \bar\nu_e$ & 0.23092 $\pm$ 0.00948 & \cr
\qquad average & 0.22376 $\pm$ 0.00259 & 0.218 -- 0.223 \cr}}$$
\hrule
\vglue 0.6truecm

\par The results on CKM matrix elements are listed in table I: the value
$\vert V_{ud}\vert =$ 0.9731(21) agrees with that from superallowed Fermi
transitions in nuclei$^{13}$, $\vert V_{ud} \vert =$ 0.9740(5), on the {\sl low
side} of the PDG ``adjusted'' range. This is because the PDG gives more weight
to the {\sl theoretical} analysis of $K_{\ell3}$ decays by Leutwyler and
Ross$^{14}$ than to the data from baryon $\beta$ decays$^{15}$; our
evaluation, based on better data (and a more complete analysis) than the
original one, but consistent with their (revised) findings$^{15}$, gives
$\vert V_{us}\vert=$ 0.2238(26) (and $\vert V_{ud}\vert^2+\vert V_{us}\vert^2
=$ 0.9970(53)), this time on the {\sl high side} of the PDG range, in accord
with expectations from CKM unitarity for three families, found valid to better
than 1 $\sigma$, and the estimate $\vert V_{ub}\vert=(4\pm2)\times10^{-3}$.

\par We next choose to sit on the low side of the PDG range for $\vert V_{ud}
\vert$ ({\sl i.e.} right on the value from Fermi transitions) and thus obtain
the values for $\vert \gamma \vert$ listed in table II. Defining a consistency
parameter $\chi^2_{cons}$ as
$$
\chi_{cons}^2 = \sum {{2 (\gamma_{rate} - \gamma_{asym})^2} \over
{\sigma_{rate}^2 + \sigma_{asym}^2}}\ ,
\eqno (13)
$$
where we assume for both $\gamma$'s the same sign, we find a definite decrease
of this parameter with $\vert V_{ud} \vert$, of more than five units over the
PDG range, to be compared with a minimum of 3.18 at our chosen value, and we
claim that {\sl all} baryon $\beta$ decays require to reduce $\vert V_{ud}
\vert$ and raise $\vert V_{us} \vert$, and advise {\sl against} using the PDG
``central values'' when precise estimates (at percent level or better) are
required. Note also that baryonic $\beta$ decay data (including superallowed
Fermi nuclear decays) have now an {\sl overall} quality much better than
$K_{\ell3}$ ones: in our opinion the error quoted by Leutwyler and Ross$^{14}$
for their estimate of $\vert V_{us}\vert$ was underestimated by at least a
factor 2 (as already advocated by Paschos and T\"urke$^{16}$).

\vglue 0.6truecm
\centerline{\bf Table II}
\centerline{\bf Axial couplings }
\vglue 0.3truecm
\hrule
$$\vbox{\halign{#\hfil&\qquad\hfil#\hfil&\qquad\hfil#\hfil&\qquad
\hfil#\hfil\cr
\qquad Decay & from rates (moduli) & from asymmetries (PDG) & SU(3) fits
\cr
$\Delta S = 0$: & & & \cr
$n \to p e \bar\nu_e$ & 1.2548 $\pm$ 0.0018 & 1.2573 $\pm$ 0.0028 & 1.2552 \cr
$\Sigma^- \to \Lambda e \bar\nu_e$ & 0.7223 $\pm$ 0.0173 & -- -- & 0.7865 \cr
$\Sigma^+ \to \Lambda \bar e \nu_e$ & 0.750 $^{+0.089}_{-0.101}$ & --  -- &
0.7865 \cr
$\Sigma^- \to \Sigma^0 e \bar\nu_e$ & -- -- & -- -- & 0.4687 \cr
$\Xi^- \to \Xi^0 e \bar\nu_e$ & $< 2 \times 10^3$ & -- -- & 0.7308 \cr
$\Delta S = 1$: & & & \cr
$\Lambda \to p e \bar\nu_e$ & 0.7251 $\pm$ 0.0112 & 0.718 $\pm$ 0.015 & 0.7308
\cr
$\Lambda \to p \mu^- \bar\nu_\mu$ & 0.756 $^{+0.128}_{-0.154}$ & -- -- &
0.7308 \cr
$\Sigma^- \to n e \bar\nu_e$ & 0.3377 $^{+0.0217}_{-0.0232}$ & -0.340 $\pm$
0.017 & -0.3178 \cr
$\Sigma^- \to n \mu^- \bar\nu_\mu$ & 0.2466 $^{+0.0664}_{-0.0928}$ & -- -- &
-0.3178 \cr
$\Xi^- \to \Lambda e \bar\nu_e$ & 0.2600 $^{+0.0411}_{-0.0490}$ & 0.25 $\pm$
0.05 & 0.2065 \cr
$\Xi^- \to \Lambda \mu^- \bar\nu_\mu$ & 0.763 $^{+0.470}_{-0.763}$ & -- -- &
0.2065 \cr
$\Xi^- \to \Sigma^0 e \bar\nu_e$ & 1.263 $^{+0.143}_{-0.161}$ & -- -- & 1.2552
\cr
$\Xi^- \to \Sigma^0 \mu^- \bar\nu_\mu$ & $< 37$ & -- -- & 1.2552 \cr
$\Xi^0 \to \Sigma^+ e \bar\nu_e$ & $< 2.76$ & -- -- & 1.2552 \cr
$\Xi^0 \to \Sigma^+ \mu^- \bar\nu_\mu$ & $< 30$ & -- -- & 1.2552 \cr}}$$
\hrule
\vglue 0.6truecm

\par The SU(3)--symmetry fit in table II leads to a $\chi^2$ of 21.98 versus
12 d.o.f., not unacceptable from a purely statistical point of view; however,
by comparing axial charges from the fit with those averaged from the data {\sl
\`a la} PDG, one can see that almost all the $\chi^2$ comes from the $\Sigma
\to \Lambda$ transition, with lesser amounts contributed by the $\Delta S = 1$
transitions other than $\Sigma^-\to n$, while of course the accurate $n \to p$
one acts as a constraint on the sum $F + D$.

\par We thus conclude that, unless the measurements on $\Sigma \to \Lambda e
\nu$ transitions are redone, yielding rates deviating upward of previous
mesurements by several standard deviations, there is solid evidence for
first--order SU(3) violation in the axial charges, which, we repeat again,
{\sl do not} show any sign of internal inconsistency warranting an increase in
their {\sl experimental} errors (if one uses state--of--the--art theoretical
analysis, not badly approximated and completely outdated formul\ae, as for
instance was done by Jaffe and Manohar$^1$).
\vglue 1.3truecm

\centerline{\bf REFERENCES}
\vglue 0.3truecm

\item{[1]} {R.L. Jaffe and A. Manohar: {\sl Nucl. Phys.} {\bf B 337} (1990)
509.}
\item{[2]} {K. Hikasa, {\sl et al.} (Particle Data Group): Phys. Rev. {\bf D
45}, N. 11, Part II (1992).}
\item{[3]} {H. Pietschmann: {\sl Weak Interactions. Formul\ae, Results and
Derivations} (Springer--Verlag, Wien 1983).}
\item{[4]} {G. Nardulli: {\sl Riv. Nuovo Cimento} {\bf 15}, N. 10 (1992).}
\item{[5]} {M. Bourquin, {\sl et al.: Z. Phys.} {\bf C 21} (1983) 27.}
\item{[6]} {P.G. Ratcliffe: {\sl Phys. Lett.} {\bf B 242} (1990) 271; M. Roos:
{\sl Phys. Lett.} {\bf B 246} (1990) 179; C. Avenarius: {\sl Phys. Lett.} {\bf
B 272} (1991) 71; A. Garcia, R. Huerta and P. Kielanowski: {\sl Phys. Rev.}
{\bf D 45} (1992) 879.}
\item{[7]} {W.J. Marciano and A. Sirlin: {\sl Phys. Rev. Lett.} {\bf 56} (1986)
22; {\sl Phys. Rev. Lett.} {\bf 61} (1988) 1815.}
\item{[8]} {W.S. Woolcock: {\sl Mod. Phys. Lett.} {\bf A 6} (1991) 2579.}
\item{[9]} {D.H. Wilkinson: {\sl Nucl. Phys.} {\bf A 377} (1982) 474.}
\item{[10]} {J.F. Donoghue and B.R. Holstein: {\sl Phys. Rev.} {\bf D 25}
(1982) 206; J.F. Donoghue, E. Golowich and B.R. Holstein: {\sl Phys. Rep.}
{\bf 131} (1986) 319; Y. Kohyama, K. Oikawa, K. Tsushima and K. Kubodera: {\sl
Phys. Lett.} {\sl B 186} (1987) 255; L.J. Carson, R.J. Oakes and C.R. Willcox:
{\sl Phys. Rev.} {\bf D 37} (1988) 3197.}
\item{[11]} {P.M. Gensini and G. Violini: Univ. Perugia report DFUPG--68--93,
subm. for publication to {\sl Nuovo Cimento A}.}
\item{[12]} {M. Ademollo and R. Gatto: {\sl Phys. Lett.} {\bf 13} (1964) 264.}
\item{[13]} {G. Rasche and W.S. Woolcock: {\sl Mod. Phys. Lett.} {\bf A 5}
(1990) 1273.}
\item{[14]} {H. Leutwyler and M. Roos: {\sl Z. Phys.} {\bf C 25} (1984) 91.}
\item{[15]} {J.-M. Gaillard and G. Sauvage: {\sl Annu. Rev. Nucl. Part Sci.}
{\bf 34} (1984) 351; revised result in ref. [2], p. III--65. See also J.F.
Donoghue, B.R. Holstein and S. Klimt: {\sl Phys. Rev.} {\bf D 35} (1987) 934.}
\item{[16]} {E.A. Paschos and U. T\"urke: {\sl Phys. Rep.} {\bf 178} (1989)
145.}

\bye